\documentclass{ws-procs975x65}

\begin{document}

\title{Gamma-ray burst host galaxies and the link to star-formation}

\author{J. P. U. Fynbo$^*$, J. Hjorth, D. Malesani, J. Sollerman, D. Watson}
\address{Dark Cosmology Centre, Copenhagen University\\
Copenhagen O, DK-2100, Denmark\\
$^*$E-mail: jfynbo@dark-cosmology.dk\\
www.dark-cosmology.dk}

\author{P. Jakobsson}
\address{Centre for Astrophysics Research, University of Hertfordshire\\
CollegeLane, Hatfield, Herts, AL109AB, UK}

\author{J. Gorosabel}
\address{Instituto de Astrof{\'i}sica de Andaluc{\'i}a (CSIC), Apartado de
Correos 3004, 18080 Granada, Spain}

\author{A. O. Jaunsen}
\address{Institute of Theoretical Astrophysics, PO Box 1029 Blindern, 0315 Oslo, Norway}

\begin{abstract}
We briefly review the current status of the study of long-duration gamma-ray
burst (GRB) host galaxies.  GRB host galaxies are mainly interesting to study
for two reasons: 1) they may help us understand where and when massive stars
were formed throughout cosmic history, and 2) the properties of host galaxies
and the localization within the hosts where GRBs are formed may give essential
clues to the precise nature of the progenitors. The main current problem is to
understand to what degree GRBs are biased tracers of star formation. If GRBs
are only formed by low-metallicity stars, then their host galaxies will not
give a representative view of where stars are formed in the Universe (at least
not a low redshifts). On the other hand, if there is no dependency on
metallicity then the nature of the host galaxies leads to the perhaps
surprising conclusion that most stars are formed in dwarf galaxies.  In order
to resolve this issue and to fully exploit the potential of GRBs as probes of
star-forming galaxies throughout the observable universe it is mandatory that a
complete sample of bursts with redshifts and host galaxy detections is built.
\end{abstract}


\bodymatter

\section{Introduction}\label{intro}

The association between long-duration (T$_{90} > 2$s\cite{chryssa93})
GRBs and massive stars and hence the link between GRBs and on-going
star-formation found its first empirical basis with the detection of the first
host galaxies (e.g.\cite{hogg99}). Subsequently, the evidence was further
strengthened with the discovery of supernovae (SNe) associated with
GRBs\cite{galama98,hjorth03,stanek03,matheson03,sollerman06,modjaz06,pian06}
(for a recent review of the GRB/SN association see\cite{woosley06}). 

Because of the link between long-duration GRBs and massive stars and due to the
fact that GRBs can be detected from both the most distant and the most dust
obscured regions in the universe GRBs were quickly identified to be very
promising tracers of star-formation (e.g.\cite{wijers98}). However, this
potential has so far not really resulted in an improved census of the
star-formation activity due to complications discussed in the next section.

A major issue currently under discussion is if GRBs are unbiased tracers of
star formation. More precisely, it is not clear if GRBs are caused by the same
(small) fraction of all dying massive stars (unbiased tracers), or if GRBs only
trace a limited segment defined by parameters such as, e.g., metallicity or 
circumstellar density (biased tracers).

The currently operating {\it Swift} satellite\cite{gehrels04} has
revolutionized GRB research with its frequent, rapid, and precise localization
of GRBs. Now it is for the first time possible in practice to use GRBs as
powerful probes. It is mandatory that this potential is exploited while {\it
Swift} is still operating.

\section{Complications in the use of GRBs as SF tracers}

\subsection{The contamination from chance projections}
\begin{figure}
\begin{center}
\psfig{file=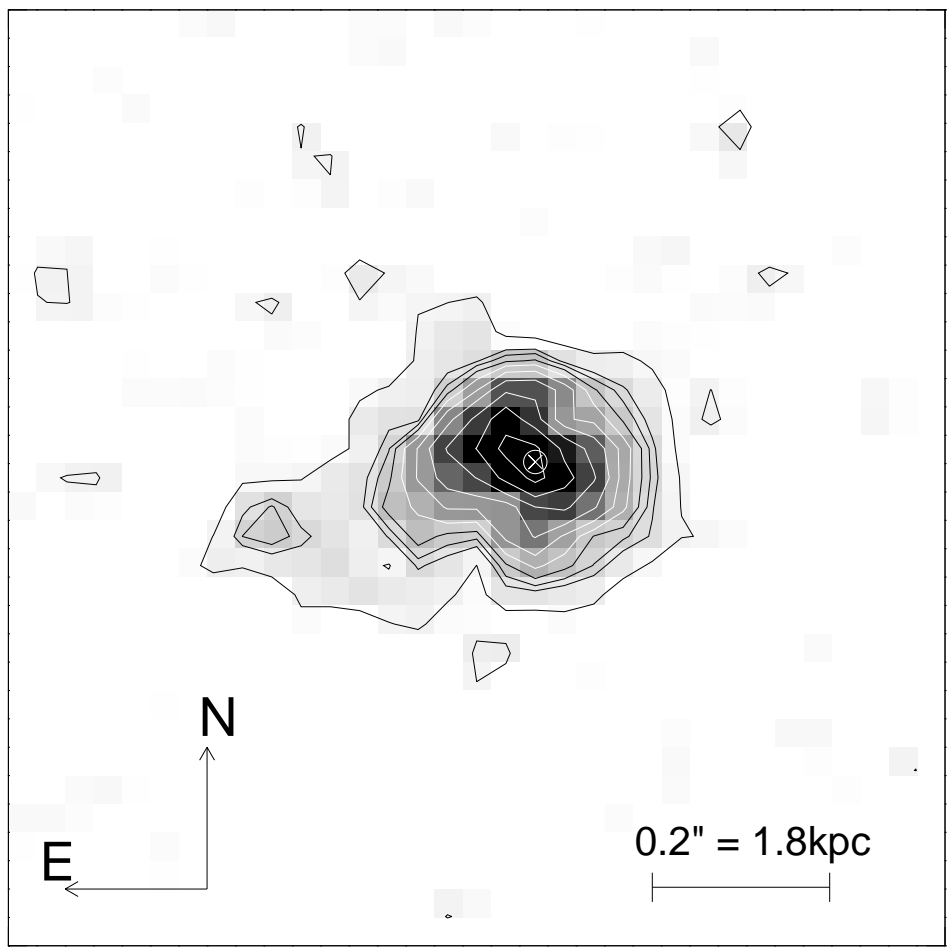,width=5in}
\caption{
The 1$\times$1 arcsec$^2$ field around the host galaxy of the $z=2.33$ HETE-II
GRB\,021004 observed with the {\it HST} (from\cite{fynbo05}).  The GRB went off
near the center of the galaxy. The position of the GRB is marked with a cross
and an error circle and in coincides with the centroid of the galaxy to within
a few hundredths of an arcsec.  In cases like this there is no problem in
identifying the correct host galaxy.
}
\label{021004host}
\end{center}
\end{figure}

The first important question to ask is: are GRB host galaxies operationally
well-defined as a class? In terms of
an operational definition the case is not so clear. If we define the host
galaxy of a particular burst to be the galaxy nearest to the line-of-sight, we
need to worry about chance projection\cite{band98}. In the majority 
of cases where an
optical afterglow has been detected and localized with subarcsecond accuracy
and where the field has been observed to deep limits a galaxy has been detected
within an impact parameter less than 1 arcsec (e.g.\cite{bloom02} and
Fig.~\ref{021004host}).  The probability for this to happen by chance is
typically found to be of the order 10$^{-3}$ or less. For a sample of a few 
hundreds GRBs chance projection should hence not be a serious concern.

When the redshift of the GRB is measured, the association can of course
be proven beyond doubt by measuring also the redshift of the potential 
host. However, the fraction of GRBs for which the redshift is measured
from the afterglow is still only $\sim$50\%\cite{palli06} so the association
cannot be made in this way for all GRBs.

In any case, there are misidentifications in the literature (see\cite{jaunsen03}
for two examples). Other noteworthy examples are the cases of GRB\,030429
and GRB\,041006: In the field of GRB\,030429 a
galaxy was detected at an impact parameter of 1.2 arcsec from the position of
the afterglow and no galaxy was detected at the position of the afterglow to
deep limits. However, spectroscopy of the afterglow and of the nearby galaxy
established that the two were unrelated\cite{palli04b}. In the case of
GRB\,041006 a galaxy 1.0 arcsec from the afterglow position was suggested
to be the host\cite{covino04}, but later {\it HST} imaging revealed another
galaxy at the afterglow position\cite{soderberg06}.

In conclusion, for GRBs localized with an accuracy worse than about 1 arcsec
chance projection cannot be excluded, but chance projection is not a major
obstacle for the use of GRBs to select star-forming galaxies. 

\subsection{Dark bursts and incomplete samples}
A crucial issue when using GRBs (or any other class of tracer) to probe
star-formation is sample selection. Whereas the detection of the GRB itself
poses no bias against dust obscured star-formation this is not the case for the
softer afterglow emission which is crucial for obtaining the precise
localization as well as measuring redshifts (see, e.g.\cite{fiore06}, and the
contribution from Dr. Fiore in these proceedings). 

In the samples of GRBs detected with satellites prior to the currently
operating {\it Swift} satellite\cite{gehrels04} the fraction of GRBs with
detected optical afterglows was only about 30\%\cite{fynbo01,lazzati02}.
Much of this
incompleteness was caused by random factors such as weather or unfortunate
celestial positions of the bursts, but some remained undetected despite both
early and deep
limits. It is possible that some of these so called ``dark bursts'' could be
caused by GRBs in very dusty environments\cite{groot98} and hence the sample of
GRBs with detected optical afterglows could very well be systematically
biased against dust obscured star formation (see also\cite{palli04,rol05} for
recent discussions of dark bursts).

In any case, such a high incompleteness imposes a large uncertainty on
statistical studies based on GRB host galaxies derived from these early 
missions.

Due to the much more precise and rapid localization capability of {\it
Swift} it is now possible to build much more complete samples (see below). 

\subsection{Are some long GRBs not associated with massive stellar death?}

Recently, it has been found that some long-duration GRBs are not associated
with SNe, namely GRB\,060505\cite{fynbo06} and
GRB\,060614\cite{fynbo06,dellavalle06,gal-yam06}. This means that either some
massive stars die without producing even faint supernovae\cite{fynbo06,
dellavalle06} or, alternatively, some long-duration GRBs are caused by
other mechanisms than collapsing massive stars\cite{gal-yam06,ofek07}. 

At least for the case of GRB\,060505 the evidence points to the former
(see\cite{fynbo06,thoene07}). However, if some long-duration GRBs indeed are
caused by other progenitors than massive stars then a new classification that
can distinguish between long GRBs from massive stars and those from other
mechanisms is required. So far no such scheme has been found
(see\cite{gehrels06}).\footnote{A classification scheme has been
proposed\cite{zhang06}, but this scheme is ambiguous and not operational (e.g.,
for GRB\,060505), involves observables that are not available for most bursts
(the associated supernova), and the nature of the progenitor, which is not
observable at all.} 

\begin{figure}
\begin{center}
\psfig{file=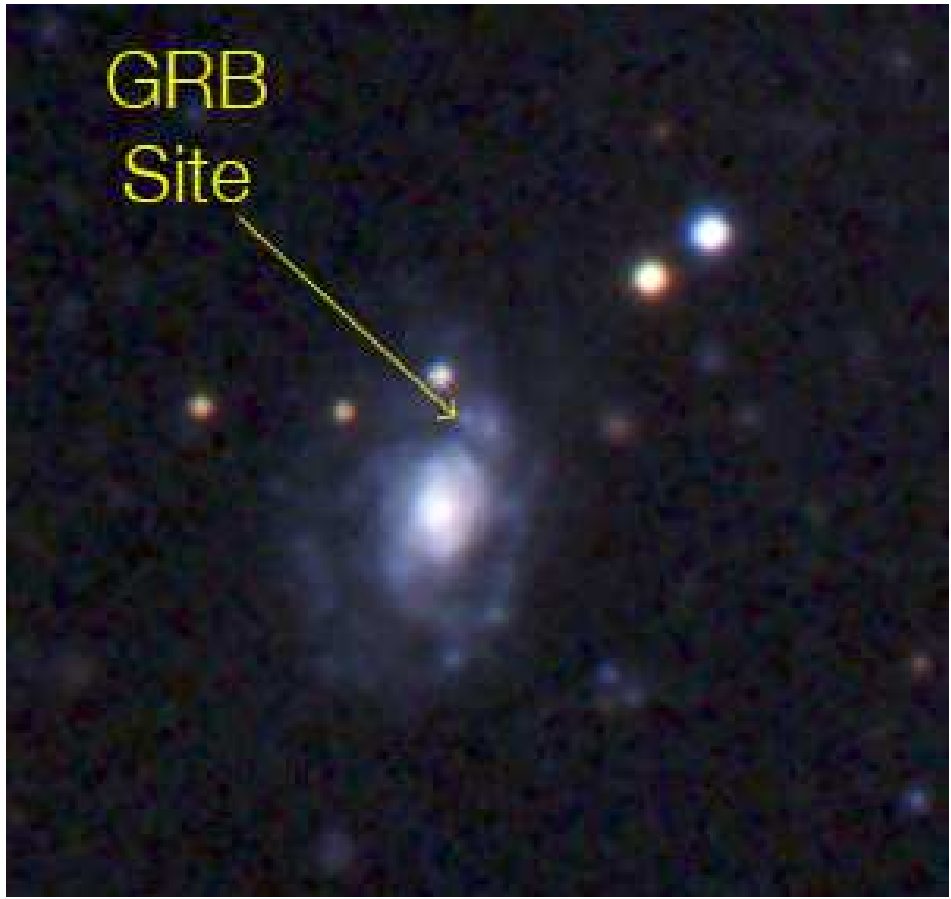,width=5in}
\caption{The host galaxy of GRB\,060505 as observed with the ESO 
VLT\cite{thoene07} (see also \cite{ofek07}.
This SN-less long GRB occurred in a
star-forming region in one of the spiral arms of a late-type host.
This is strong evidence that the progenitor was a massive star.
The properties of this host is also within the range found for 
other long GRBs. As an example, the host galaxy of GRB\,050824 is
very similar in terms of luminosity and star-formation rate\cite{sollerman07},
and the location within the host is similar to the location of other
long GRBs in spiral hosts\cite{fynbo00,sollerman02,lefloch02,palli05b}.
}
\label{0060505host}
\end{center}
\end{figure}

\section{What do GRB host galaxies tell us about star formation 
despite the complications?}

\subsection{Early results}
Despite the complications mentioned above, the previous decade of GRB host
galaxy studies has after all taught us a lot about GRBs and star formation
(see \cite{vanparadijs00,djorgovski03} for early reviews).
GRB hosts were early on found to be predominantly faint, blue star-forming
galaxies\cite{hogg99}. The early studies found that these properties of GRB
hosts were consistent with the expectation if GRBs are unbiased tracers of
star-formation\cite{mao98,hogg99}. It was also realized early on that GRBs
offer a unique possibility to locate and study star-formation activity in 
dwarf galaxies at $z>2$\cite{jensen01}. This is basically impossible with
any other currently existing method. The star-formation rates were found to
be modest, but the specific star-formation rates are among the highest ever
found\cite{christensen04}. GRB hosts are hence typically in a starburst
state.

\subsection{Current status}
As the samples have grown larger it has become clear that GRB hosts
in general have 
low luminosities. As an example, not a single GRB at $z>2.5$ has
a host galaxy that is bright enough that it would be detected in the 
ground based surveys for Lyman-break galaxies at similar redshifts (i.e.
R$<$25.5). This implies that either the Lyman-break galaxies are just the tip
of an iceberg of star-forming galaxies at $z>2.5$ or GRB are biased 
towards faint hosts. Also nearby GRB hosts are found to be of low
luminosity\cite{sollerman05,stanek06}.

The environments of GRB host galaxies has not been subject of many studies.  At
low redshifts \cite{foley05} studied the field of the host galaxy of
GRB\,980425, that was reported to be member of a group. However, based on
redshift measurements of the proposed group members, \cite{foley05} could
establish that the host of GRB\,980425 is an isolated dwarf galaxy. The host
galaxy of GRB\,030329 was found to have a companion at projected distance of
600 kpc\cite{gorosabel05}. The host galaxy of GRB\,060505 has been located to
the outskirts of a filament extending from the galaxy cluster Abell 
3837\cite{thoene07}.
At redshifts $z\gtrsim$2 a few GRB fields have been
studied using narrow band Ly$\alpha$ imaging \cite{fynbo02, fynbo03, palli05}.
In all cases several other galaxies at the same redshift as the GRB host were
identified, but it is not sure whether the galaxy densities in these fields are
higher than in blank fields as no blank field studies have been carried out at
similar redshifts. However, the density of Ly$\alpha$ emitters were found to be
as high as in the fields around powerful radio sources that have been proposed
to be forming proto-clusters, which would suggest that GRBs could reside in
overdense fields at $z\gtrsim$2. Another study\cite{bornancini04} argue
for a low galaxy density in GRB host galaxy environments, but this study
is based on data probing very small fields-of-view around the hosts.

The absence of bright dust-obscured hosts in the GRB host samples (but see
also\cite{berger03}) has also been
argued to be inconsistent with a scenario where GRBs are unbiased tracers
\cite{lefloch03,tanvir04}, but, as it has recently been pointed
out\cite{priddey06}, ``there is sufficient uncertainty in models and
underlying assumptions, as yet poorly constrained by observation (e.g. the
adopted dust temperature) that a correlation between massive, dust-enshrouded
star formation and GRB production cannot be firmly ruled out.''

In addition, GRB hosts have been found to
be more frequent Ly$\alpha$ emitters than Lyman-break galaxies at similar
redshifts, which could be due to a low metallicity
preference\cite{fynbo03}. However, all of these studies targeted very
incomplete pre-{\it Swift} samples and this raises the question whether
the faint hosts can be explained by, e.g., a bias against dust obscured
GRB afterglows. 

A very important result is that GRBs and core-collapse SNe are found in
different environments\cite{fruchter06}. The same study also found that GRB
host galaxies at $z<1$ are fainter than the host galaxies of core-collapse SNe.
This study is also based on incomplete pre-{\it Swift} samples, but as the SNe
samples are also biased against dusty regions this result does seem to be
strong evidence that GRBs are biased towards massive stars with low metallicity
(see also\cite{wolf06,modjaz07}). A study of GRB host galaxy morphology 
based on nearly the same data as in \cite{fruchter06} has, however, concluded
that "GRB hosts trace the starburst population at high redshift, as similarly
concentrated galaxies at $z>2$ are undergoing a disproportionate amount of star
formation for their luminosities. Furthermore, our results show that GRBs are
not only an effective tracer of star formation but are perhaps ideal tracers of
typical galaxies undergoing star formation at any epoch, making them perhaps
our best hope of locating the earliest galaxies at $z>7$"\cite{conselice05}.  A
similar conclusion has been reached in other studies\cite{fynbo06b,alma07}.

A recent study of a sample of {\it Swift} GRBs have found that {\it Swift}
GRBs appear to have fainter hosts than pre-{\it Swift} GRBs\cite{ovaldsen},
presumably due their larger distance\cite{palli06}.

\section{Progenitor models}
GRB progenitor models also give some indication of whether GRBs will be biased
tracers. In the collapsar models for a GRB the central ingredients are
formation of a rapidly rotating black hole and removal of the Hydrogen/Helium
envelope before the explosion (see\cite{woosley06,fryer07}). The need for rapid
rotation suggests that the progenitors need to have a low metallicity stellar
wind, in order to avoid removing angular momentum from the star. On the other
hand, the need to remove the Hydrogen envelope suggests that a high metallicity
is needed at least for some single star progenitor models (e.g.\cite{heger03},
their Fig.~7. See also\cite{fryer05}). Binary progenitor models have also been
explored\cite{fryer05}. We note that low metallicity massive stars in the
Magellanic clouds do not appear to rotate faster than massive stars in the
Milky Way despite their lower metallicities\cite{penny04}.  There is also
evidence that some GRBs are not related to the formation of a black hole, but
rather a neutron star\cite{mazzali06,maeda07}. There is also evidence for
sustained activity of the inner engine\cite{grupe06,sollerman07}. Whether the
currently favored models are consistent with this evidence is unclear to the
authors.  Therefore, there is still some open issues concerning the role of
metallicity in GRB progenitor models. There are other models for long duration
GRBs where there is no obvious dependency on metallicity\cite{dar04,ouyed05}. 

\section{Outlook}

A number of steps need to be taken before the potential of GRB host galaxies
as probes of the cosmic star-formation activity can really be exploited.

\subsection{Building a complete sample}
A crucial obstacle to overcome 
is to collect a complete sample and to secure as high a redshift completeness
as possible. As pointed out by Dr. Jakobsson in these proceedings, efforts are
ongoing to this end. The sample includes only {\it Swift} bursts fulfilling:

\begin{itemize}
\item{T$_{90} > 2$ s.}
\item{A rapidly distributed XRT error circle.}
\item{Low foreground Galactic extinction: A$_\mathrm{V} < 0.5$ mag.}
\item{Declinations within [-70,70] degrees.}
\end{itemize}

We are currently working on detecting the hosts and measuring redshifts for as
many of these bursts as possible. By December 2006, there is a total of 85 of
these bursts, 43 with spectroscopic redshifts measurements (with a mean value
of $z_{med}=2.41$), 22 with no afterglow detection, and 20 with optical and/or
IR afterglow detection but unknown redshift either because no spectrum was
secured or no lines were detected on which to base a redshift.

\subsection{Getting Dark bursts into the picture}
\begin{figure}
\begin{center}   
\psfig{file=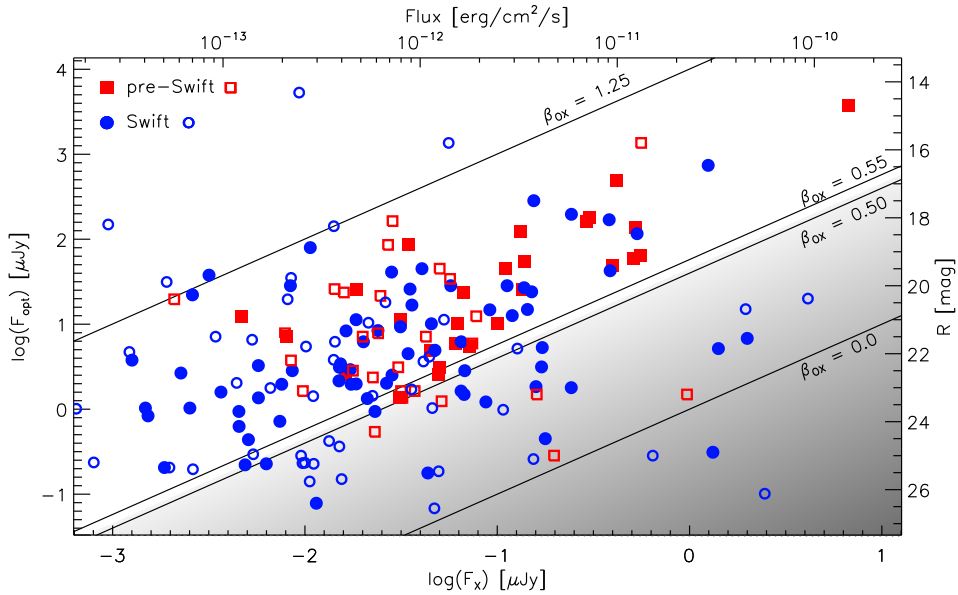,width=5in}
\caption{
A diagram of optical flux vs. X-ray flux for all long-duration
bursts with available data (until 1 Jan 2007). The magnitudes
have been corrected for Galactic extinction. Filled symbols
indicate optical detections while open symbols are upper limits.
Lines of constant optical-to-Xray spectral slope 
$\beta_{\mathrm{OX}}$ are shown along with
the corresponding value. We define dark bursts as those which
have $\beta_{\mathrm{OX}} < 0.5$. The \emph{Swift} dark bursts
fraction is around 25$\%$ compared to the pre-\emph{Swift}
value of 10$\%$. Note that the majority of the \emph{Swift}
X-ray fluxes have been obtained from GCNs which only give an
approximate value. Adapted from \cite{palli04}.
}
\label{darkbursts}
\end{center}
\end{figure}

Another crucial obstacle is to tackle the question of dark bursts. In
Fig.~\ref{darkbursts} we show an updated version of the dark burst figure from
\cite{palli04}. For {\it Swift} GRBs at least 25\% are dark\footnote{Note that
a dark burst here is defined as a burst with an X-ray to optical slope too
shallow for
the synchrotron fireball model.  Hence, according to this definition, a burst
with no detected optical afterglow is not necessarily a dark burst, and some
dark bursts have detected optical afterglows (GRB~050401 is an 
example\cite{watson06}).}.  An
important step forward has been taken with the refined astrometry purely based
on the X-ray afterglows\cite{butler06}. Now a substantial fraction of GRBs only
detected in the X-rays have astrometry accurate to be better than 2 arcsec.
This still leaves some probability for chance alignment, but it is now possible
to make statistical comparison of X-ray based error-circles for dark bursts and
bursts that are not dark and look, e.g., for a higher density of
dusty galaxies in the dark burst error circles.

\subsection{Simulations}
Finally, there is a need for more predictions of which properties we expect of
GRB host galaxies under different assumptions about the progenitors of GRBs.
Could the fact that most GRB hosts are so faint simply be telling us that this
is the type of galaxies responsible for the majority of the star-formation
activity? At redshifts $z\gtrsim2$ the faint end slope for star-forming
galaxies has been found to be very steep so this is not an unreasonable
suggestion\cite{fynbo02,palli05}. There has so far only been a few such studies
trying to predict the properties of GRB hosts (e.g.\cite{courty04, nuza06,
courty07}),
but much more work is needed (e.g. using larger simulated volumes and
exploring different recipes for including star-formation in the simulations).

\end{document}